# Malware Classification using Deep Neural Networks: Performance Evaluation and Applications in Edge Devices

**Akhil M R[1], Adithya Krishna V Sharma[2], Harivardhan Swamy[1], Pavan A[1], Ashray Shetty[1], Anirudh B Sathyanarayana[1]**

[1]Student, Dept. of Computer Science and Engineering, PES University, Bengaluru, Karnataka, India
[2]Associate Software Engineer, Red Hat, Bengaluru, Karnataka, India

**Abstract -** *With the increasing extent of malware attacks in the present day along with the difficulty in detecting modern malware, it is necessary to evaluate the effectiveness and performance of Deep Neural Networks (DNNs) for malware classification. Multiple DNN architectures can be designed and trained to detect and classify malware binaries. Results demonstrate the potential of DNNs in accurately classifying malware with high accuracy rates observed across different malware types. Additionally, the feasibility of deploying these DNN models on edge devices to enable real-time classification, particularly in resource-constrained scenarios proves to be integral to large IoT systems. By optimizing model architectures and leveraging edge computing capabilities, the proposed methodologies achieve efficient performance even with limited resources. This study contributes to advancing malware detection techniques and emphasizes the significance of integrating cybersecurity measures for the early detection of malware and further preventing the adverse effects caused by such attacks. Optimal considerations regarding the distribution of security tasks to edge devices are addressed to ensure that the integrity and availability of large-scale IoT systems are not compromised due to malware attacks, advocating for a more resilient and secure digital ecosystem.*

*Key Words*: **Cybersecurity, Data Protection, Deep Neural Networks, IoT Security, Malware Classification, Performance Evaluation**

## 1. INTRODUCTION

Combating the constant spread of malware continues to be a major concern in the constantly changing world of cybersecurity. The integrity, confidentiality, and availability of digital assets are seriously threatened by malicious software, or malware, making accurate malware categorization a critical component of contemporary security systems. Long used for malware detection, traditional signature-based methods are ineffective against fast-evolving and zero-day malware. This calls for the adoption of more advanced methodologies. Deep neural networks (DNNs), a particularly noteworthy development in deep learning in recent years, have shown significant promise in several areas, including image identification, natural language processing, and autonomous systems. In the field of malware categorization, its capacity to automatically learn sophisticated patterns and features from raw data has attracted interest. DNNs can dramatically improve the precision and effectiveness of malware detection by sifting through malware samples and deriving useful representations.

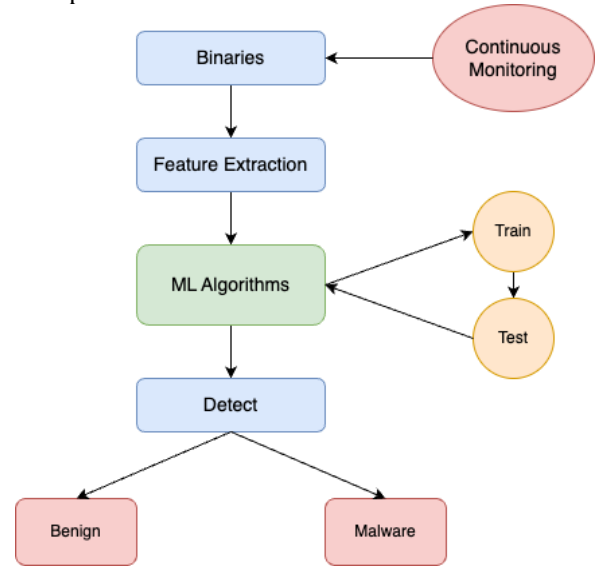

**Fig–1**: ML based Malware Classification Flow

IoT devices are complex in nature and are subject to a wide variety of cyber-attacks with malware attacks being one of the prominent ones. Additionally with the increasing adoption of IoT devices in industries, these IoT systems will experience a rise in cyber-attacks. Therefore, it is deemed necessary to deploy efficient methodologies to detect and mitigate the adverse effects which would otherwise be caused by such malware attacks. According to Quoc-Dung Ngo et al., IoT malware detection techniques can be broadly classified into two domains, namely static analysis, and dynamic analysis [5].

Dynamic analysis includes having to execute the binaries and monitor for any malicious activity which could potentially infect the real time execution environment. In contrast, static analysis involves analyzing the binaries without executing them. [5] The methodologies explored in this paper leverage deep learning techniques to identify patterns and classify malware binaries without having to execute them.

Additionally, we cover a crucial topic of implementing advanced malware classification algorithms in contexts with limited resources. The computational and memory resources of edge devices, such as Internet of Things (IoT) gadgets and low-powered computer systems, are constrained. For effective and real-time malware detection at the network edge, it is critical to assess the applicability of our deep neural network approach in such devices. Therefore, the computation time or the latency to classify malware binaries is measured once the trained model is obtained. This research intends to advance cybersecurity procedures by examining the functionality and applicability of our DNN-based malware classification methodology. By providing security experts with a cutting-edge tool for malware detection that is early and accurate, our research has the potential to increase the resilience of digital ecosystems to the ever-increasing cyberthreats.

## 2. RELATED WORK

Previous research on malware classification can be broadly categorized into two main approaches: they are,

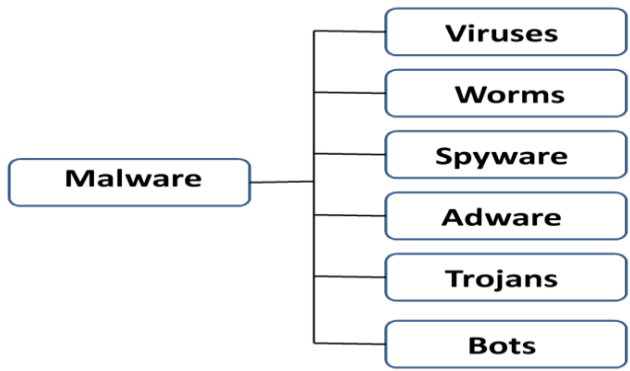

**Fig –2**: Types of Malwares

### 2.1 Non-Machine Learning Models

Traditionally, malware detection relied on non-machine learning techniques, such as static or dynamic signature-based methods. Static analysis involves examining the syntax or structural properties of the program to identify malware before its execution. However, malware developers employ various encryption, polymorphism, and obfuscation techniques to evade these detection algorithms. In the dynamic approach, malware is executed in a controlled virtual environment, and its behavior is analyzed to detect harmful actions during or after execution. While dynamic analysis shows promise, it remains complex and time-consuming. The major drawback of classical signature-based detection is its lack of scalability, and its effectiveness can be compromised with the emergence of new variants of malware. As a result, researchers have turned to intelligent machine learning algorithms as an alternative approach.

**Dynamic Analysis:** Researchers have made significant efforts to propose behavior-based malware detection methods that capture program behavior at runtime. One approach is to monitor the program's interactions with the operating system through the analysis of API calls. To develop effective and robust systems, some studies consider additional semantic information, such as the sequence of API calls and the use of graph representations. These approaches analyze the temporal order of API calls, the effect of API calls on registers, or extract behavioral graphs based on dependencies between API call parameters. In contrast to program-centric approaches, global, system-wide methods have been proposed, such as an access activity model by Lanzi et al. [8], which captures generalized interactions of benign applications with operating system resources, resulting in a low false positive rate. However, dynamic analysis techniques face challenges in handling execution-driven datasets, security precautions during experimentation, and dynamic anti-analysis defenses used by modern malware to evade detection.

**Static Analysis:** On the other hand, static approaches perform analysis without executing the program. The research literature demonstrates a wide variety of static analysis methods, with SAFE [11] and SAVE [10] being influential heuristic static malware detection approaches. These works proposed using different patterns to detect malicious content in executable files. Since then, numerous techniques have emerged based on different malware attributes, such as the header or body of the Portable Executable (PE) file, with analysis conducted on bytecode or by disassembling the code to extract opcodes and other relevant information. The main challenge in static analysis is coping with packing and obfuscation. Recently, generic approaches for the automatic de-obfuscation of obfuscated programs have been proposed. Additionally, static techniques have been employed to assess if a detected malware is like a previously seen variant without performing costly unpacking.

### 2.2 Machine Learning Models

To address the limitations of non-machine learning methods and capitalize on the shared behavior patterns among malware variants, anti-malware organizations have developed sophisticated classification methods based on data mining and machine learning techniques. These methods employ various feature extraction methods to build intelligent malware detection systems, often using SVM-based classifiers, Naïve Bayes classifiers, or multiple classifiers [9].

For example, Nataraj et al. [7] proposes a strategy to represent malware as grayscale images and use GIST to compute texture features, which are then classified using a k-nearest neighbor algorithm. However, these shallow learning techniques suffer from scalability issues with the growing number of malware samples and require manual feature engineering. To overcome these challenges, the current research focuses on developing deep learning architectures that are more robust and applicable to various malware samples.

While some techniques target superior performance on specific datasets, like the Microsoft Malware Dataset [12], we aim to construct a more versatile framework applicable to any type of malware sample. For instance, Drew et al. [13], [14] employed a modern gene sequence classification tool for malware classification on the Microsoft Malware Dataset. Ahmadi et al. [15] trained a classifier based on the XGBoost technique, while the winning team of the Microsoft Malware Classification Challenge (BIG 2015) utilized a complex combination of features with the XGBoost classifier.

Another related work proposed in [16] involves the application of a CNN for malware classification. The author experimented with three different architectures by adding an extra block consisting of a convolutional layer followed by a Max-pooling layer each time to the base model. However, their model remains relatively shallow. In contrast, our research delves into exploring deeper CNN architectures for improved malware classification.

## 3. DATA AVAILABILITY AND PREPARATION

For the purpose of demonstrating the effectiveness of DNNs on malware binaries, the dataset chosen was MaleVis [6]. The MaleVis [6] dataset contains 14,226 malware images spanning across 26 classes which also includes 1 cleanware class. From the dataset, 10 malware classes were sampled and a total of 1400 images were further sampled from these classes overall for the purpose of training.

For testing and validation purposes, a total of 550 images were sampled spanning across the 10 classes. The images in the MaleVis [6] dataset was obtained by extracting the binary images from the malware files in 3 channel RGB format. The images are then resized into square sized resolutions of 224x224 and 300x300.

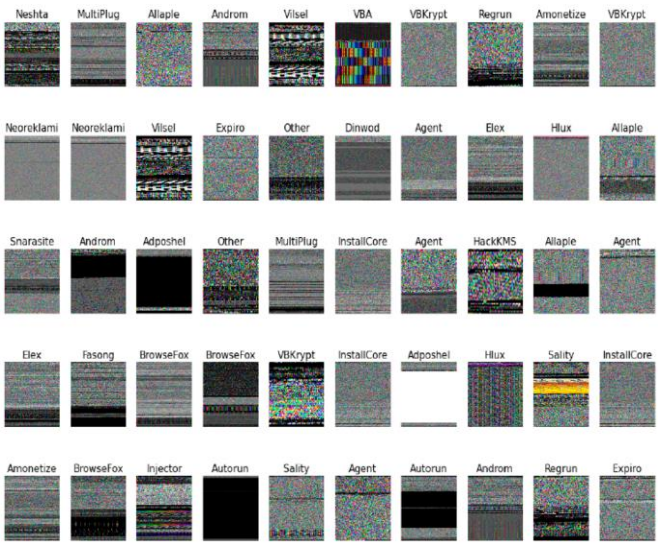

**Fig -3:** Images pertaining to the classes of the MaleVis dataset [6]

**Table -1:** Classes Sampled for the Purpose of Training

| Class ID | Family | Malware Category | Sample Size |
|---|---|---|---|
| 1 | Adposhel | Adware | 140 |
| 2 | Agent | Trojan | 140 |
| 3 | Allaple | Worm | 140 |
| 4 | Amonetize | Adware | 140 |
| 5 | Androm | Backdoor | 140 |
| 6 | Autorun | Worm | 140 |
| 7 | BrowseFox | Adware | 140 |
| 8 | Dinwod | Trojan | 140 |
| 9 | Hex | Trojan | 140 |
| 10 | Expiro | Virus | 140 |

## 4. IMPLEMENTATION METHODOLOGY

The experimental setup involved training the models for 10 epochs on a system with RTX 3050 as the GPU. For increasing the effectiveness of the models, pre-trained imagenet weights were imported and applied before initiating the training process. A learning rate of $10^{-4}$ was used while also being configured to be adaptive in nature during the training process with the minimum allowed learning rate being $10^{-7}$.

We run extensive tests to gauge the precision and effectiveness of our method in order to evaluate its performance. We do this by comparing the deep neural network's classification accuracy against unseen samples after training it on a broad array of malware samples. One of the key metrics used in the evaluation of resource efficiency is computational latency. This computational latency was measured as the time taken to classify the set of 550 test images. Other metrics such as accuracy, recall and F1 score were also taken into consideration while testing the model and are covered below. The details pertaining to the models utilized are explored as follows.

### 4.1 ResNetV2

Deep convolutional neural networks (CNNs) present issues, therefore ResNetV2 is an extension of ResNet created to address those challenges. By introducing "bottleneck" blocks that compress feature maps, it can retain efficiency while lowering computational complexity. To reduce deterioration and hasten convergence during training, "pre-activation" modules place batch normalization and ReLU activation before convolutions. ResNetV2 performs better than its predecessor, especially in more complex network topologies, displaying increased training effectiveness and precision. ResNetV2, a pioneering architecture in computer vision research, has been widely used for image classification, object recognition, and semantic segmentation applications. Its breakthroughs advance state-of-the-art in image recognition applications by solving gradient problems and optimizing learning functions in deep CNNs.

### 4.2 DenseNet201

DenseNet201 is a deep convolutional neural network architecture that extends the DenseNet concept by employing 201 layers. It utilizes dense blocks, where each layer receives feature maps from preceding layers, facilitating feature reuse, and mitigating the vanishing gradient problem. This densely connected structure fosters efficient information flow and parameter sharing, resulting in improved memory utilization and better gradient propagation during training. With its substantial depth, DenseNet201 excels in learning complex patterns and representations from data, making it highly effective for various computer vision tasks such as image classification, object detection, and semantic segmentation. Its exceptional performance on benchmark datasets has solidified DenseNet201 as a leading architecture in the field of deep learning for visual recognition tasks.

## 4.3 InceptionNetV3

An improved convolutional neural network architecture called InceptionNetV3, sometimes known as Inception V3, was created for image identification applications. It provides numerous parallel convolutional layers of various filter sizes to effectively capture features at various scales and resolutions, building on the strengths of its forerunners, InceptionNet and Inception V2. In order to capture both fine-grained and global characteristics, the "Inception module" concurrently uses 1x1, 3x3, and 5x5 convolutions. Meanwhile, "Factorized 7x7" convolutions lessen computational complexity without sacrificing the receptive field.

With the use of batch normalization and auxiliary classifiers, it also improves convergence and addresses the vanishing gradient issue. Global average pooling minimizes the number of parameters and avoids overfitting. InceptionNetV3 has been widely used for research and practical applications because of its exceptional performance in picture classification, object identification, and visual recognition tasks.

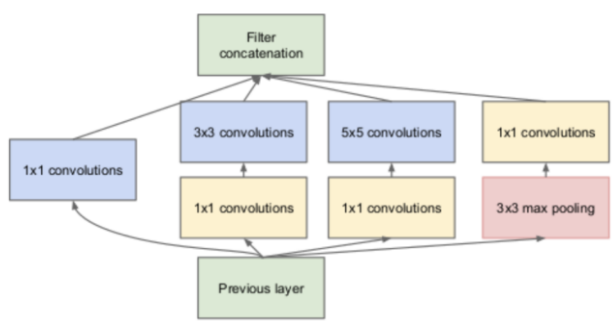

**Fig -4:** MobileNet Architecture

## 4.4 Xception

The deep convolutional neural network architecture known as Xception, short for "Extreme Inception," was unveiled by Google and was motivated by the Inception idea. It uses "depth wise separable convolutions," which combine depth wise and pointwise convolutions, to replace conventional standard convolutions while maintaining accuracy.

Xception speeds up training and inference times by improving feature learning and parameter efficiency, making it the best choice for computer vision workloads, especially in contexts with limited resources like mobile devices and edge computing. Xception has established itself as a leading deep learning model and a popular option for image recognition applications thanks to its outstanding performance.

## 4.5 MobileNet Small

A variation of the MobileNet architecture called MobileNet Small is designed for quick and effective deep learning on devices with limited resources. It significantly decreases the model size and computational complexity by using depth wise separable convolutions, ensuring excellent performance on mobile devices and embedded systems.

In spite of its effectiveness, MobileNet Small retains respectable accuracy in jobs like object detection and image categorization. It is an ideal option for on-device AI applications because of this design decision, which enables real-time processing and reduces computational and energy expenses. MobileNet Small, which is widely used in edge computing applications, demonstrates its usefulness in enhancing deep learning for mobile and embedded devices.

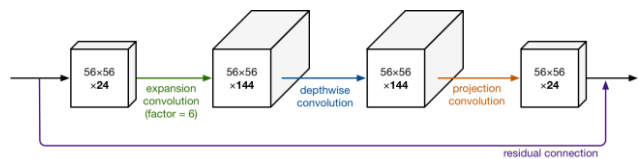

**Fig -5:** MobileNet Architecture

## 4.5 MobileNet Large

It is a lightweight deep learning architecture designed for efficient image classification on mobile devices. It also utilizes depth wise separable convolutions, a width multiplier, and a resolution multiplier to reduce computational complexity and model size. Despite its efficiency-focused design, MobileNet Large maintains competitive accuracy and is well-suited for real-time applications on resource-constrained devices, making it a significant advancement in the field of computer vision.

In summary, MobileNet Small sacrifices some accuracy for even greater efficiency and compactness, making it ideal for scenarios where minimizing model size and computational requirements are critical, while MobileNet Large strikes a balance between efficiency and accuracy, making it more suitable for general-purpose mobile vision applications on devices with moderate resources.

## 5. RESULTS AND DISCUSSIONS

**Table -2:** Results Obtained from Training Various DNNs

| Model | Compute Latency | Accuracy | Recall | F1 Score |
|---|---|---|---|---|
| ResNetV2 | 8.062 | 86.54 | 86.27 | 86.78 |
| DenseNet201 | 10.87 | 94.54 | 94.43 | 94.42 |
| InceptionNetV3 | 8.33 | 91.81 | 91.53 | 91.64 |
| Xception | 8.11 | 93.63 | 93.68 | 93.64 |
| MobileNet-Small | 3.51 | 85.63 | 84.52 | 81.77 |
| MobileNet-Large | 6.17 | 88.01 | 87.92 | 87.87 |

The above table summarizes the results obtained from testing various DNNs on the MaleVis [6] dataset. The compute latency depicts the time taken in seconds to classify 550 test images sampled from the dataset. It was observed that DenseNet201 achieved the highest accuracy in comparison to the other models during the test run, although a tradeoff between the computational latency and accuracy can be significantly noticed.

DenseNet201 showed the highest latency to compute along with an increased model accuracy. MobileNet-small on the other hand showed an accuracy on par with that of ResNetV2 with an exceptional computational latency of just **3.51 seconds**.

This proves that with effective fine tuning of the model, it could be deployed viably in real-world scenarios as well. MobileNet-large showed exceptional results achieving an accuracy higher than that of its smaller counterpart version, but with a slight tradeoff with the computational latency.

Furthermore, the above set of results can be utilized for choosing the right model for deployment in resource constrained scenarios as per the requirement and the availability of computational power in edge devices.

## 6. CONCLUSION

In this survey article, we have explored the application of deep neural networks (DNNs) for malware classification. Malware detection and classification are critical tasks in today's cybersecurity landscape due to the ever-evolving nature of malicious threats. Traditional non-machine learning methods such as static and dynamic analysis have been widely used but are facing challenges in coping with the increasing complexity and diversity of malware.

The machine learning methods section focused on DNN architectures, namely ResNet, DenseNet, InceptionNet, Xception, MobileNet Small, and MobileNet Large. These DNNs have demonstrated promising results in various computer vision tasks and have shown potential for tackling malware classification as well.

From the performance evaluation, it is evident that DNN architectures can effectively detect and classify malware binaries with high accuracy and improved generalization. DenseNet201 showed the best performance among the models evaluated with an accuracy of 94.5. The ability to handle large-scale datasets and learn intricate patterns allows DNNs to discern even the most sophisticated malware variants. Moreover, transfer learning techniques can be leveraged to adapt pre-trained models on related tasks, reducing the data requirements and training time.

Regarding the applicability in edge devices, the compact nature of some DNNs like MobileNet Small and MobileNet Large allows for efficient deployment on resource-constrained devices, such as IoT devices and smartphones. The ability to perform classification on the edge can enhance real-time threat detection and response, mitigating the need for constant cloud communication and reducing latency.

However, societal concerns also need to be addressed when using DNNs for malware classification. There are ethical and privacy considerations related to data collection, model fairness, and potential misuse of these technologies. It is crucial to adhere to robust privacy policies and ensure the transparency and accountability of the deployed models.